# *The vapor controller effect on physical properties of CdTe thin films*

M.H. EHSANI, H. REZAGHOLIPOUR DIZAJI, M.H. MIRHAJ
*Thin Film Lab., Physics Department, Semnan University, Semnan, I.R.Iran*

*Abstract:*
This paper reports the optical and structural properties of 400 nm nanocrystalline cadmium telluride thin films fabricated using a new vapor controller system in vacuum. This system is indeed a rotating cylinder placed in the material steam path during coating process. The upper cross section of the cylinder has one arc groove allowing the steam to pass through. Three type samples were made under different vapor controlling conditions namely; without controller system (conventional method), with rotating cylinder, and with upper cross section of the cylinder alone which acts as a rotating shutter. The prepared samples have been studied by X-ray diffraction (XRD), Field Emission Scanning Electron Microscopy (FESEM) and UV-Vis Spectroscopy techniques. The results show, not only the structural properties were improved by this technique but also the band gap of samples were changed.

## 1- Introduction

CdTe is a direct-band gap semiconductor with high optical absorption coefficient ($>10^4 cm^{-1}$). It is used as the photon absorber in thin-film solar cells bearing their names. More than 90% of the incident light is absorbed in a few micrometers of the material without a need to phonon assisted mechanism [1]. CdTe is commonly paired with CdS to form the heterojunction that separates the charge carriers produced by photon absorption. The CdS layer is essential for electronic junction formation, but the incident photons on solar cell pass through CdS (window layer) and absorbed in the CdTe (absorber layer). Bonnet and Rabenhorst [2] were those who first reported an efficiency of 6% for CdTe/CdS thin film solar cells. It took nearly 30 years that a research group of NREL reported a record efficiency of 16.5%, which was 65% of the theoretical value [3]. In order to gain the maximum efficiency, a lot of research work has been already reported on CdTe/CdS thin films solar cells which are focused on three main areas, namely; cell structure and fabrication, cell modeling and accelerated life testing, and characterization of materials and devices [4-12].

The effects due to grain size have attracted the attention of many researchers who are engaged in the area of materials characterization. This is because, the grain boundaries are generally considered to act as strong electrical recombination centers, barriers to current transport, and are also known as the main source for considerable leakage currents even one regent in light trapping [1, 13-14]. Therefore, minimizing the impact of grain boundaries by maximizing the grain size has been the main aim of many researchers to gain higher device performance. On the other hand, using the nano-crystalline n-type CdTe (small grain size) as a window layer in n-Nano-CdTe/p-bulk CdTe homostructures is another attractive subject [15].

*There are various methods in order to control the grain size such as heat treatment in vacuum or in the atmosphere of air, oxygen, or CdCl$_2$ [13, 16-18]. Major and coworkers reported the use of different nitrogen pressures during the deposition process for the purpose of controlling the grain size [19]. A series of research activities has been carried out by Patil and coworkers to control the grain size of variety of thin film materials.*

*They have used the vapor chopping technique to prepare thin films of tin oxide, bismuth oxide, and magnesium oxide and some other materials by vacuum thermal evaporation method [20-26]. Vapor chopping provides more time for the evaporated ad-atoms to settle on the substrate than the continuous arrival of evaporated ad-atoms on the nonchopped thin film. The chopper they used was a V-shaped (angle 155°) circular metallic vane placed between substrate and boat. The chopper speed was ~5–6 rot/s which provided growth flux interruptions at a constant rate. They have reported the increase in packing density, adhesion, refractive index and decease in intrinsic stress and transmission loss of thin films due to the use of vapor chopping technique. A decrease in the number of voids due to lateral surface diffusion of depositing particles in the film was obtained due to vapor chopping.*

*In the present work, for controlling the grain size of CdTe thin films, a new vapor controlling system was employed to deposit CdTe thin films under different conditions. That is, rotating cylinder, and upper cross section of the cylinder alone (rotating shutter). The results were compared with those of CdTe thin film prepared by conventional thermal evaporation technique. The prepared films were studied using XRD, FESEM, UV-Vis measurements.*

## 2-New vapor controlling system design

*Fig. 1 shows a cylinder with the capability of rotation around its axes through which it is coupled to a dc motor. The motor is fed by a dc power supply from outside of the chamber. The cylinder consists of a disk at its upper side which acts as a shutter. There is an arc groove in the disk with the depth equal to the disk thickness and length of 10 cm. The cylinder height, diameter and sheet thickness are 130mm, 100mm and 2mm respectively. All the parts are made of stainless still. This set-up provides the possibility to investigate the effect of a quasi closed space system usage on the physical properties of thin films*

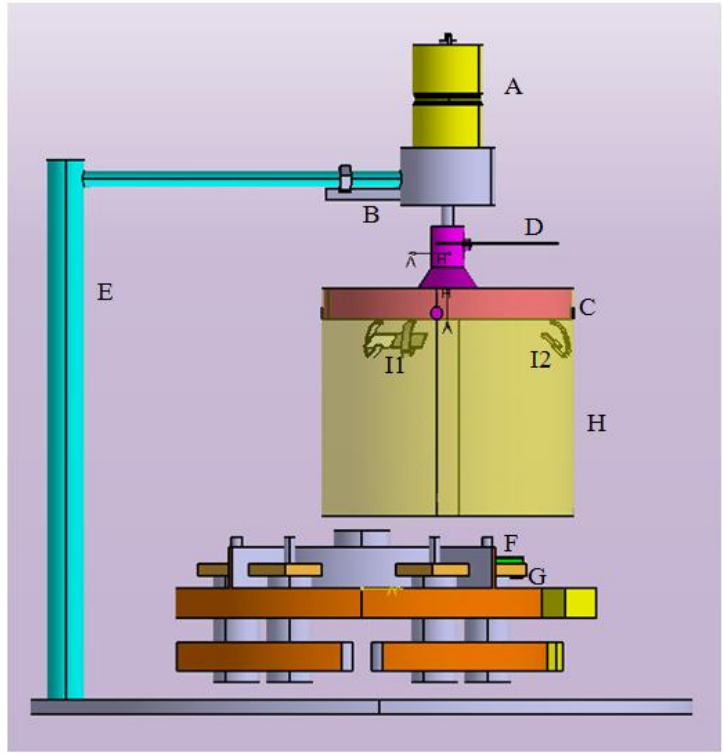

*Fig. 1: schematic of a new vapor controller system ; (A) dc motor, (B) motor holder, (C) revolving disk(shutter), (D) substrate, (E) stand, (F) boat, (G) electrode ,(H) cylinder, (I1 I2) two extra substrate holders*

*When a direct bias voltage is applied to the dc motor, the disk revolves and the vapor way towards the substrate is opened and closed periodically. The vapor flow of material is condensed on the substrate surface after passing through the groove. When the vapor is not allowed to reach the surface, the atoms on the surface will have some times to migrate and find more suitable place on it to settle. Fig. 2 shows the designed vapor controlling system without the cylinder, i.e. rotating shutter. The disk can be rotated clockwise and anticlockwise at different rates using the dc motor. The period for a complete rotation of the disk can be controlled by changing the applied voltage. The fixed shutter (B) shown in the figure is used to prevent the contamination of the other places in the chamber during coating process.*

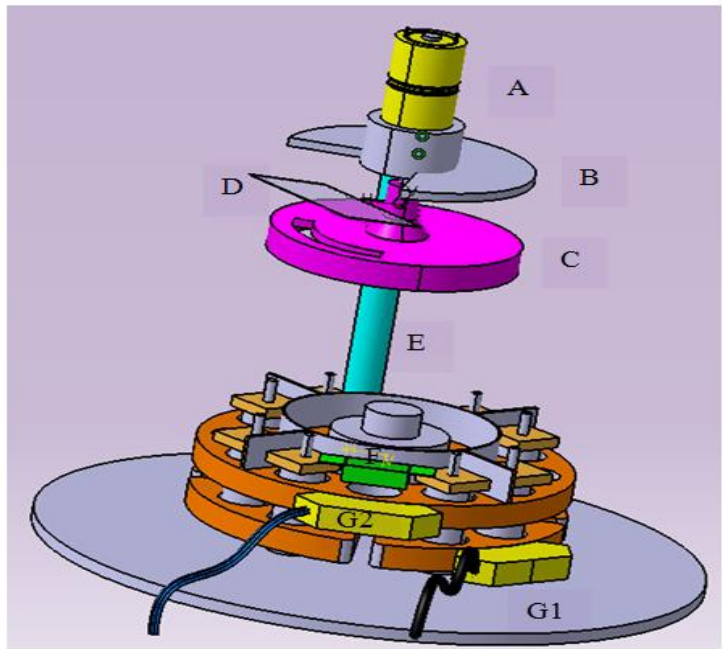

*Fig. 2: schematic of vapor controlling system consisting; (A) dc motor, (B) fixed shutter, (C) revolving disk, (D) substrate, (E) holder shaft, (F) boat, (G1 and G2) the two electrodes ,(H) arc groove*

*The designed vapor controller system has the following advantages:*
*-The possibility of investigation on the effect of vapor flow path geometry by changing the groove shape, angular velocity of the rotating cylinder, etc.*
*-The possibility of mounting the substrates at different angels inside the closed system.*
*-Less contamination of the inner body of the chamber during the coating process.*

## 3-Experiment procedure

*CdTe films of 400 nm thick were deposited on glass substrates in a HindHiVac coating unit (Model15F6) under different conditions, namely; by conventional method, with rotating cylinder and with rotating shutter. The substrates were cleaned in acetone and methanol using ultrasonic bath and then dried by nitrogen gas. Further, the substrates were subjected to glow discharge cleaning before deposition. The base pressure of the chamber was $10^{-6}$ mbar. CdTe powder of 99.99% purity supplied by Aldrich Company was evaporated from a capped molybdenum boat. This kind of boat was used in order to avoid the CdTe powder sputtering out during the deposition process. The deposition rate was measured and controlled in situ using HindHiVac thickness monitor (Model DTM-101). The typical growth rate was 5Å/s on average. The source-to-substrate distance was approximately 15cm. All the films were deposited on substrate maintained at ambient temperature in vacuum chamber.*

CdTe thin films were deposited on glass substrate by conventional thermal evaporation method and by new vapor controlling system (rotating shutter and rotating cylinder).When using the rotating cylinder, it was mounted in such a way as to direct the vapor flow of whole the vaporized material towards the substrate only. The rotation speed of the cylinder was 5 rpm. The same rotation speed was applied when the rotating shutter was used.After deposition under any one of the above conditions; the films were removed from the coating chamber and exposed to the ambient atmosphere.

The crystal structure of CdTe films was determined by x-ray diffraction (XRD) using an x-ray diffractometer (Advance Model D8) with high intensity $Cuk_\alpha$ radiation ($\lambda=1.5406Å$). The optical transmission spectrum of the prepared CdTe thin layers was obtained in Shimadzu UV-Vis spectrophotometer (Model UV-1650 PC). The surface of the films was examined by Field Emission Scanning Electron Microscopy (FESEM).

## 4-Results and discussion
### 4-1 Structural studies

Fig. 3 demonstrates the comparison among the XRD patterns of CdTe thin films made by the three different arrangements. All the three specimens showed zincblende structure with [111] preferred orientation.

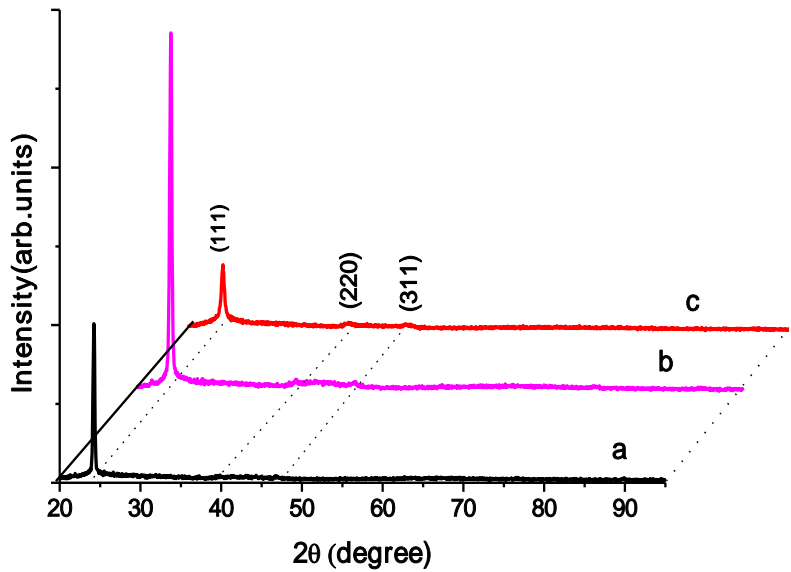

*Fig.3: The XRD patterns of CdTe thin films made by (a) conventional method, (b) rotating shutter and (c) rotating cylinder*

Employing the rotating shutter caused the peak due to (111) plane to intensify to about three times compared to that of the film made by conventional method and also to intensify the peaks due to (220) and (311) planes little more. The increased intensity can be attributed to the improved crystallinity of the grown films. It is seen from this figure that, when a rotating cylinder is used, the (111) peak intensity reduces and the other two peaks are still stronger than those appearing on the XRD pattern of the film made by conventional method.

Fig. 4 provides a more exact look at the less intense parts on the three XRD patterns shown in Fig.3. As it is observed, two more peaks due to (440) and (511) planes appeared on the XRD pattern of the CdTe film made by rotating shutter system.

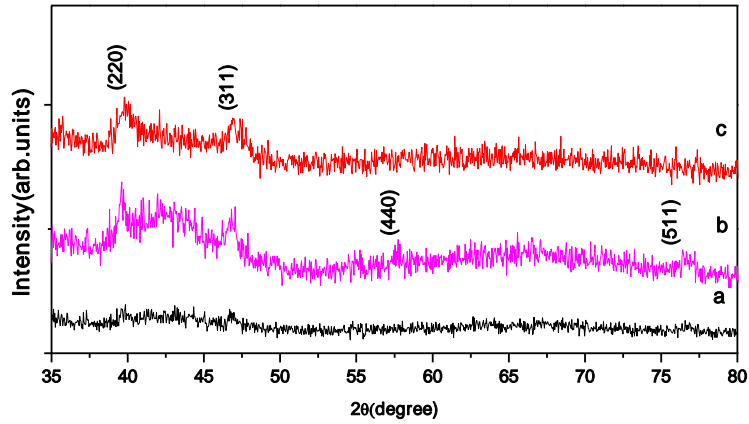

Fig. 4: scaled up less intense parts of XRD patterns shown in Fig.3

This may be attributed to the relaxation time given to the atoms condensed on the surface before the other atoms arrived at the film surface due to the rotating shutter usage. Hence the vapor chopped films were found to be more crystalline than the nonchopped one which was also reported by Tamboli et al [27]. They suggested that the structural improvement could be attributed to the increment in the film density.

The crystallite sizes of CdTe thin films prepared at three different vapor flow conditions in [111] direction were measured using Scherrer's formula.

$$D = 0.9\lambda / (\beta \cos\theta) \qquad (1)$$

Where $D$ is the average crystallite size, $\lambda$ is the X-ray wavelength equal to 1.5406Å, $\theta$ is the Bragg angle and $\beta$ is the corrected full width at half maximum [FWHM]. The crystallite size was evaluated to be ~32nm, ~26nm and ~19nm for the films made by conventional thermal evaporation, rotating shutter and rotating cylinder methods respectively. It may be interpreted that some kind of stirring effect on the films' surface may happen due to employing this vapor controller system which in turn, can prevent the growth of large crystallites. Patil et al also

*observed that the crystallite size of vapor chopped bismuth oxide films was smaller than that of nonchopped films [24].*

### *4-2 Film surface studies*

*The surface of the three specimens was studied by FESEM. It can be said from a comparison among the three figures 5 (a –c) that the grain size of the CdTe thin film made by conventional method is bigger than that of the films made by the invented methods. It is also seen that the smoothness of thin film increases due to chopping. The same observation has been reported by Tamboli et al [27]. The films made by the invented methods show more uniform surface morphology as compared to the one made by the conventional method which indicates that the thin film uniformity increases upon chopping.*

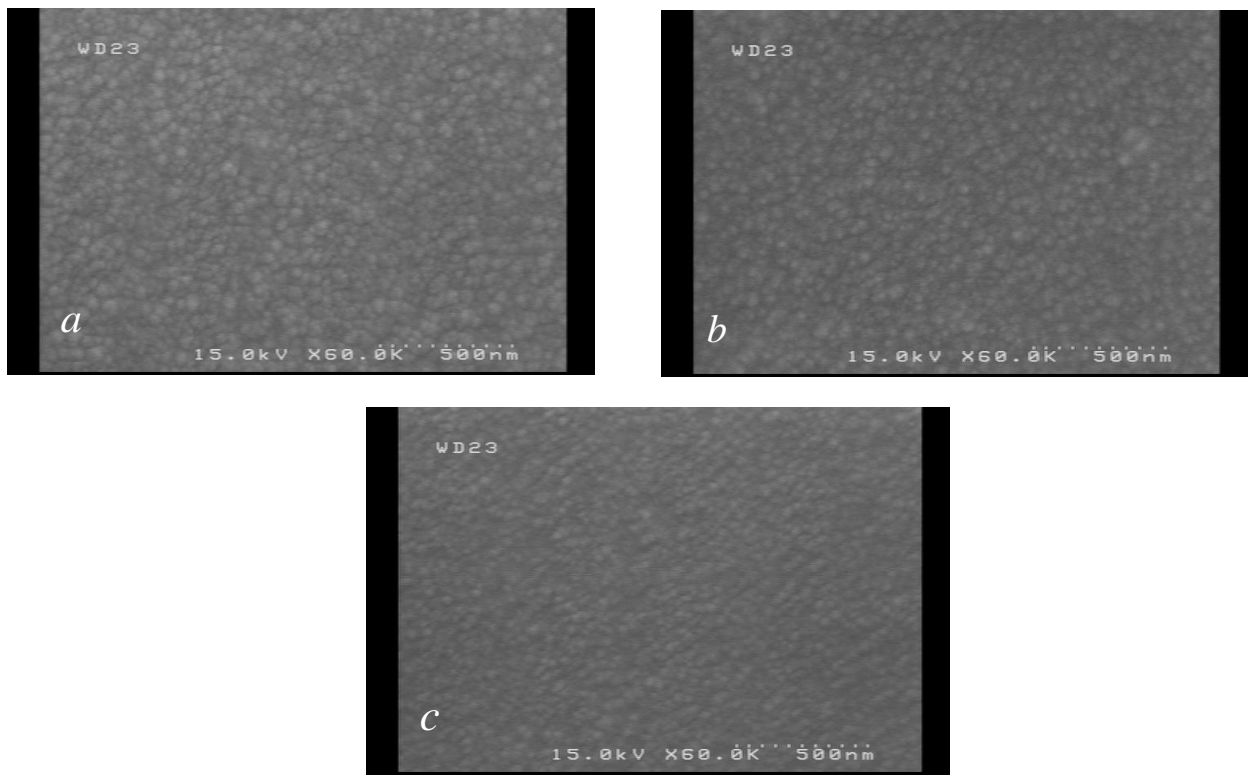

*Fig. 5: the surface morphology of the CdTe thin films made by (a) conventional, (b) rotating shutter system and (c) rotating cylinder methods*

## 4-3 Optical studies

Fig. 6 shows the absorbance of the CdTe thin films made by the three methods. It is observed that the specimen made by the conventional method has higher absorbance compared to the films prepared by the invented methods. Therefore, it may be deduced that the CdTe films made by new vapor controlling system with lower absorbance may be a good candidate for use as a window layer in solar cells.

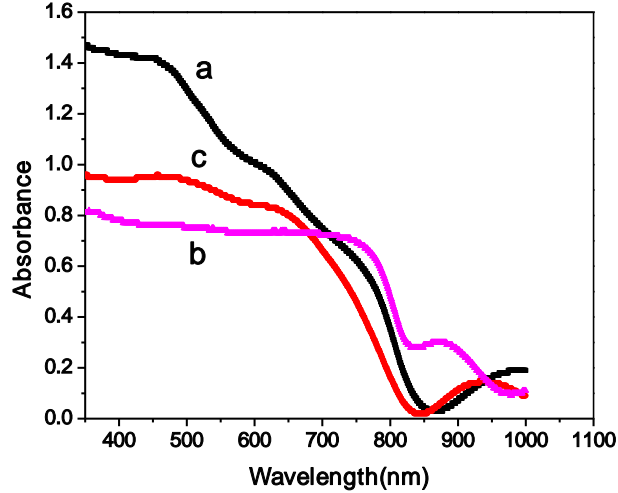

*Fig. 6: Optical absorbance spectra of the surface morphology of the CdTe thin films made by (a) conventional, (b) rotating shutter and (c) rotating cylinder methods*

From the absorbance data, the absorption coefficient α was calculated using Lamert Law [28]:
$$Ln(I_0/I)=2.303A=\alpha d \qquad (2)$$
Where $I_0$ and $I$ are the incident and transmitted light intensity respectively, A the optical absorbance and d the film thickness. The absorption coefficient, α, was found using the following relation [28].
$$\alpha= [A_0 (h\nu-E_g)^{1/2}]/h\nu \qquad (3)$$
Where $A_0$ is a constant related to the effective masses associated with the bands and $E_g$ is the band gap energy.

Figure 7 gives the plots of $(\alpha h\nu)^2$ versus $h\nu$ for the CdTe thin films made by the three methods which has been drawn using Eq.(3). Extrapolating the linear portion of the curves to the x-axis (α=0) will give the band gap.

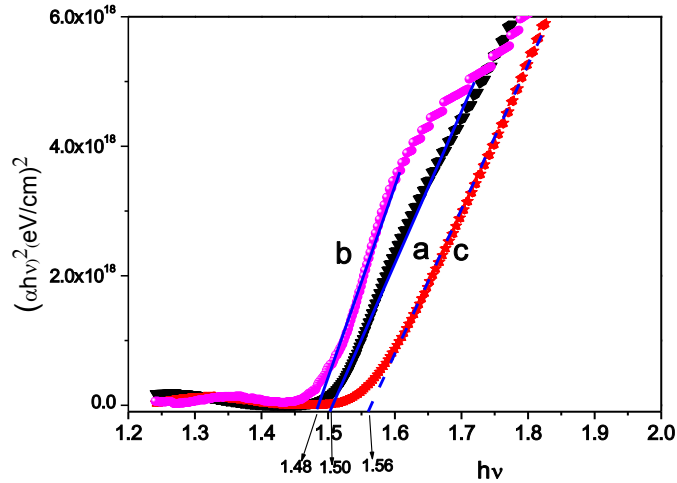

*Fig. 7: Graphs of (αhν)² Vs hν for CdTe thin films made by (a) conventional, (b) rotating shutter and (c) rotating cylinder methods*

*The $E_g$ value for the CdTe thin films made by conventional, rotating shutter system and rotating cylinder methods was found to be 1.50 eV, 1.48 eV and 1.56 eV respectively. The obtained values show that the absorption edge of the film deposited with rotating shutter is the lowest one and close to the band gap of single crystal CdTe thin film, i.e. 1.42 eV. This can be attributed to the improvement in the films crystallinity as supported by the XRD studies. The bang gap increase is related to the grain size effect. Considering the Eg value obtained by employing the rotating cylinder method, one may attribute it to the reduction in the film grain size. Similar behavior have been observed by Rusu and coworkers when they deposited CdTe thin films using a close space system by vacuum thermal evaporation technique[29-30].*

## Conclusion

*Nano structured CdTe thin films were deposited on glass substrate by conventional thermal evaporation method and by a new vapor controlling system (rotating shutter and rotating cylinder).The XRD results showed that this new system could decrease the grain size of CdTe thin films. This was also confirmed by the FESEM photographs. The designed vapor controller system affected the optical properties through changing the optical band gap. The sample made by the help of rotating shutter showed a band gap near to that of CdTe single crystal. The main effect of employing the rotating cylinder is its capabilities for decreasing the grain size.*